\documentclass[letterpaper, 10pt, conference]{ieeeconf}

\usepackage[utf8]{inputenc}
\usepackage[T1]{fontenc}
\usepackage{graphicx}
\usepackage{makecell}
\usepackage[most]{tcolorbox}
\usepackage{color, colortbl}
\usepackage{soul}
\usepackage{subcaption}
\definecolor{Gray}{gray}{0.9}
\definecolor{LightGray}{gray}{0.6}

\IEEEoverridecommandlockouts   
\newcommand{\lakmal}[1]{\textcolor{black}{#1}}

\title{\LARGE \bf
Multimodal Earable Sensing for Human Energy Expenditure Estimation}

\author{Yasith Amarasinghe$^{\dagger}$$^{*}$, Darshana Sandaruwan$^{\dagger}$$^{*}$, Thilina Madusanka$^{\dagger}$$^{*}$, Indika Perera$^{\dagger}$ and Lakmal Meegahapola$^{\ddagger}$
\thanks{$^{\dagger}$Yasith Amarasinghe, Darshana Sandaruwan, Thilina Madusanka, and Indika Perera are with the University of Moratuwa, Sri Lanka}
\thanks{$^{*}$Equal contribution.}
\thanks{$^{\ddagger}$Lakmal Meegahapola is with Idiap Research Institute \& EPFL, Switzerland.}
}

\begin{document}

\maketitle

\begin{abstract}
Energy Expenditure Estimation (EEE) is vital for maintaining weight, managing chronic diseases, achieving fitness goals, and improving overall health and well-being. Gold standard measurements for energy expenditure are expensive and time-consuming, hence limiting utility and adoption. Prior work has used wearable sensors for EEE as a workaround. Moreover, earables (ear-worn sensing devices such as earbuds) have recently emerged as a sub-category of wearables with unique characteristics (i.e., small form factor, high adoption) and positioning on the human body (i.e., robust to motion, high stability, facing thin skin), opening up a novel sensing opportunity. However, earables with multimodal sensors have rarely been used for EEE, with data collected in multiple activity types. Further, it is unknown how earable sensors perform compared to standard wearable sensors worn on other body positions. In this study, using a publicly available dataset gathered from 17 participants, we evaluate the EEE performance using multimodal sensors of earable devices to show that an MAE of 0.5 MET (RMSE = 0.67) can be achieved. Furthermore, we compare the EEE performance of three commercial wearable devices with the earable, demonstrating competitive performance of earables. 

\textit{Clinical Relevance} -- This study confirms that multimodal sensors in earables could be used for EEE with comparable performance to other commercial wearables.
\newline

\end{abstract}

\section{INTRODUCTION}
Energy Expenditure Estimation (EEE) is crucial in understanding and managing chronic diseases like obesity, diabetes, and metabolic disorders \cite{ee_survey_paper}. In addition, monitoring energy expenditure (EE) can assist individuals with fatigue and sleep disorders \cite{eee_sleep_disorder}. A deeper understanding of an individual's EE is an important factor that facilitates the creation of personalized treatment plans for conditions such as obesity \cite{martinez}. Further, the accurate tracking of EE also enables individuals to gain insight into their physical activity, energy consumption, and net calorie intake \cite{lagerros}. This empowers them to make informed decisions and manage their weight by maintaining a calorie deficit or surplus, depending on their goals. Furthermore, in sports, measuring EE is essential for developing better training plans, monitoring performance, and preventing injuries like spinal cord injuries \cite{price}.

The Total Energy Expenditure (TEE) {comprises} three main components: Physical Activity Energy Expenditure (PAEE), Diet Induced Thermogenesis, and Basal Metabolic Rate. Among these components, PAEE is considered the most significant to quantify, and our study focuses on this component specifically. For the purpose of our study, from here on we will refer to PAEE as EE.

There are three main methods for EEE that are considered gold standards but have their own limitations \cite{ee_survey_paper}. These methods are: \textit{ i)} Direct Calorimetry, considered to be the most expensive method, \textit{ii)} Indirect Calorimetry, which has practical issues, and \textit{ iii)} Doubly Labeled Water, which cannot provide any means of near real-time EE readings. However, all these methods are costly, need specialized equipment, and are not well-suited for large-scale use. Further, the Metabolic Equivalent Task (MET) is the unit that is usually used for measuring EEE \cite{ee_survey_paper}, hence used in this study. But some studies have used the unit kilocalories per minute (kcal/min) for EEE as well \cite{ee_survey_paper, altini, rothney}.

Wearable devices such as fitness trackers and smartwatches, are widely used in health tracking \cite{health_wearable_devices}. They are equipped with sensors that can measure activity, heart rate, and sleep patterns, among other aspects. Furthermore, wearable devices are also used for measuring EE, and they overcome many limitations in gold standard techniques \cite{ee_survey_paper}. These devices have been used in one or more body locations for EEE in the past literature. For example, Rothney et al. \cite{rothney} achieved decent results (MAE of 0.29 kcal/min) for EEE using accelerometers placed on the hip, while Cvetkovi{\'c} et al. \cite{cvetkovic} achieved MAE of 0.76\;MET using three smartphone accelerometer sensors in the trouser pocket, the bag, and on the chest. Although some studies have already used multiple types of  sensors, such as heart rate in combination with  accelerometer sensors \cite{altini}, it is important to check whether different sensor combinations can improve EEE.

Due to their specific positioning on the human body, earables offer a distinct platform for sensing a wide range of aspects relevant to health \cite{roddiger}. The thin skin of ears improves the accuracy of the measurements of embedded sensors like photoplethysmography (PPG) \cite{colvonen2021response}. According to Röddiger et al. \cite{roddiger}, studies have proven that earables can achieve medical-grade heart rate accuracy with PPG sensors in resting conditions. Further, earable devices are portable and small in size, which allows them to be worn comfortably throughout the day. When considering EEE using earables, Bouarfa et al. \cite{bouarfa} used an ear-worn device embedded with a research-grade single accelerometer sensor and got an MAE of 1.2\;MET. They measured EE against ten different activities. Furthermore, LeBoeuf et al. \cite{leboeuf} also studied an earbud-based solution to measure EE. The study used a combination of two sensors inside their earbuds: an accelerometer and a PPG sensor. However, they conducted EEE only against cardiopulmonary exercise. 

%In conclusion, earables have shown potential in measuring various physiological parameters and EEE.

Hence, to the best of our knowledge, no studies used earables with multimodal sensors, including an accelerometer, gyroscope, and PPG, for EEE across multiple activity types and different intensity levels. Further, no study compared the EEE performance of earables with other wearable devices. Therefore, we formulated the following two research questions: \textbf{RQ1}: Can earables with multimodal sensors be used for EEE, with data collected during multiple activity types and levels? \textbf{RQ2}: How does the performance of earable sensors compare to other commercial devices placed on different body positions?

By addressing the above research questions, this paper makes the following contributions.

\begin{itemize}
    \item Using a novel, publicly available multidevice and multimodal sensor dataset collected from 17 participants during various levels of physical activity (from sedentary to vigorous), we showed that earable devices could be used for EEE with reasonable performance. We obtained an MAE of 0.5 (RMSE = 0.67) for regression using an XGBoost model, on par with previous literature.
    \item We provided a detailed comparison between earable and other wearable devices for EEE. The results showed that earables have comparably decent results, just behind Zephyr BioHarness (chest-belt) and surpassing Empatica E4 (wristband) and Muse S (headband). Hence, we believe with their positioning and stability on the human body, earables have the potential to surpass or match the capabilities for EEE of other wearable devices.
\end{itemize}
\section{DATASET DESCRIPTION}

\paragraph{Dataset}

\lakmal{A novel and publicly available multidevice and multimodal dataset collected from 17 participants (12 males and 5 females) was used for this study \cite{weee}. Data were captured during the execution of three specific activities: resting, cycling, and running.} Participants were provided with eight different wearable devices including an Indirect Calorimeter device for the ground-truth measurements. In addition to the sensor data, the dataset includes demographic and body composition details, activity details, and data collected through questionnaires given to each participant. Although eight devices have been used in the data collection, only five devices were selected for this study due to inconsistencies and missing data. The devices and contained sensors were \footnote{We use the following terminology for brevity: Accelerometer -- ACC, Gyroscope -- GYRO, Photoplethysmography -- PPG, Electrodermal Activity -- EDA, Temperature -- TEMP, Electroencephalography -- EEG, Electrocardiography -- ECG, Breathing Rate -- BR.}: \textit{i)} Nokia Bell Labs Earbuds (NBL): ACC, GYRO, PPG; \textit{ii)} Empatica E4 Wristband (EE4): ACC, PPG, EDA, TEMP; \textit{iii)} Muse S Headband (MSH): ACC, GYRO, EEG; \textit{iv)} Zephyr Bioharness (ZBH): ACC, ECG, BR; and \textit{v)} VO2 Master Analyzer Face Mask: VO2 (ml/kg/min) as the gold standard ground truth.

\iffalse 
\begin{table}[t]
    \centering
    \caption{sensor data of each device used for the study}
    \label{tab:dataset_description}
    \begin{tabular}{ll}
         \rowcolor[HTML]{EDEDED}
         \textbf{Device} &
         \textbf{Sensor Data Used}
         \\

        \arrayrulecolor{Gray}
        
         Nokia Bell Labs Earbuds (NBL) &
         \begin{tabular}{@{}l@{}}
            Accelerometer\\
            Gyroscope\\
            Photoplethysmography
         \end{tabular}
         \\

         \arrayrulecolor{Gray}
         \hline

        Empatica E4 (EE4) &
        \begin{tabular}{@{}l@{}}
            Accelerometer\\
            Photoplethysmography\\
            Electrodermal Activity\\
            Temperature
        \end{tabular}
        \\

        \arrayrulecolor{Gray}
        \hline

        Muse S headband (MSH) &
        \begin{tabular}{@{}l@{}}
            Accelerometer\\
            Gyroscope\\
            Electroencephalography
        \end{tabular}
        \\

        \arrayrulecolor{Gray}
        \hline
        
        Zephyr BioHarness (ZBH) &
        \begin{tabular}{@{}l@{}}
            Summary of sensor data\\
            created per each participant
        \end{tabular}
         
        \\

        \arrayrulecolor{Gray}
        \hline
        
        VO2 Master Analyzer face mask &
        VO2 (ml/kg/min)
        \\
        
        \arrayrulecolor{Gray}
        \hline
    \end{tabular}
\end{table}
\fi 

\paragraph{Study Protocol}

\begin{figure}[t]
    \centering
    \hspace*{-2mm}
    \includegraphics[scale=0.12]{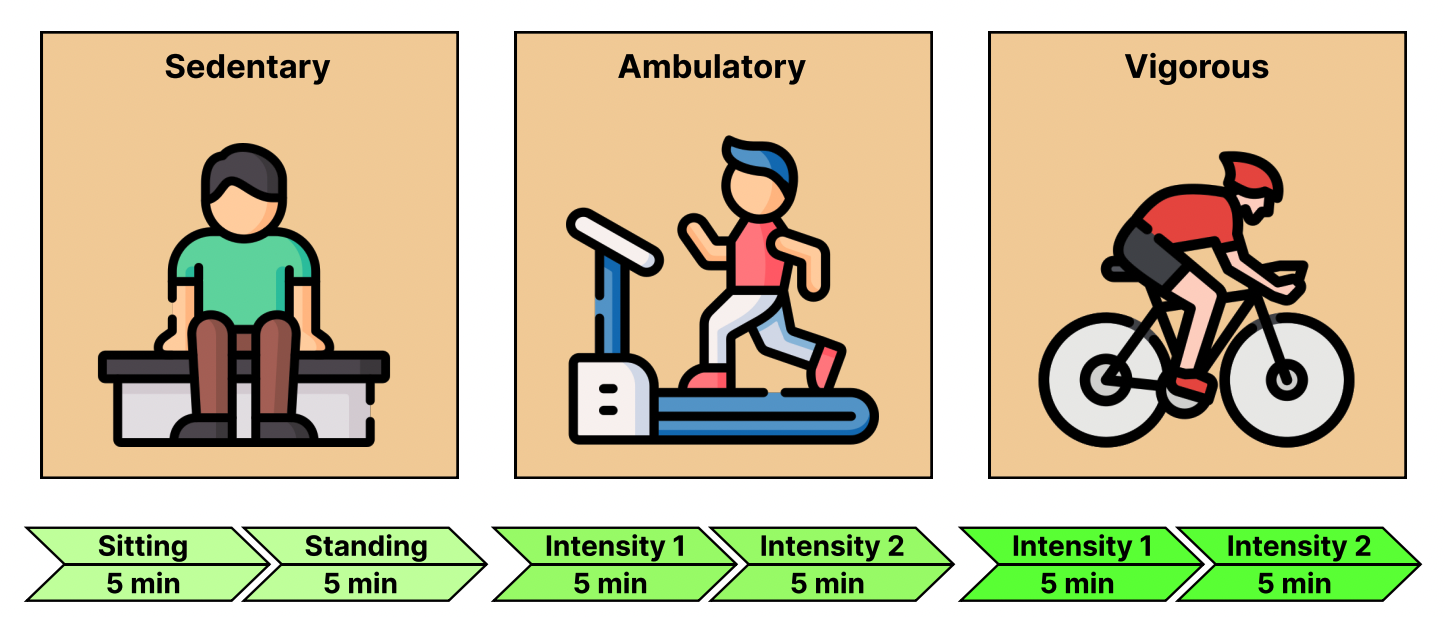}
    \caption{Summary of the study protocol \cite{weee}.}
    \label{fig:weee_study}
    \vspace{-0.1 in}
\end{figure}

The study included three main activities. Each activity was done under two intensity levels to replicate real-life scenarios. Each activity was conducted for 10 minutes; 5 minutes with the first intensity level and 5 minutes with the second intensity level. Fig. \ref{fig:weee_study} provides information about the activities carried out.

\section{METHODOLOGY}

\paragraph{Pre-Processing}

Some devices include similar types of sensor modalities. For example, accelerometer sensors are embedded in Nokia Bell Labs earbuds, Empatica E4 wristband, and Muse S headband. But these sensor data were in different units. First, we converted those units into a common unit \lakmal{(e.g., Accelerometer sensor data were in three different units \footnote{The unit "g" stands for "gravity". i.e., the acceleration due to gravity on the Earth's surface is approximately 9.8 ms\textsuperscript{-2}, or 1g.}; EE4: g/64, NBL - milli g, MSH - g. Therefore, the first two units were converted into g)}. To reduce the high-frequency noise of the accelerometer sensor data of these devices, an overlapping central moving average filter with a window of one minute was used \cite{weee}. PPG data were available on Empatica E4 and Nokia Bell Labs earbud devices. They were processed using HeartPy \cite{heartpy} toolkit. A 3\textsuperscript{rd} order Butterworth Band Pass filter was applied to remove the noise of the PPG data with cutoff frequencies as 0.7\;Hz (42\;bpm) and 3.5\;Hz (210\;bpm) \lakmal{\cite{heartpy}}. After cleaning the data, a 10-second central moving average was used to extract various types of heart-related data. EDA data were available under the Empatica E4 wristband, and features were extracted using NeuroKit \cite{neurokit}. MET values were then calculated according to the equation (\textit{1\;MET = 3.5\;ml\;O$_{2}$\;/\;kg\;/\;min}) as suggested in literature \cite{hills2014assessment}, using ground-truth Indirect Calorimeter (VO2) data. Then, a one-minute time window was dropped from the start and the end of each activity to remove any noisy data that could emerge during the change of activities. Then a dataset per each device, per each time window (i.e., 2, 4, 6, 8, 10, and 12-second time windows was used. Although prior work on EEE suggests 4 to 10 seconds time windows are ideal for EEE \cite{ee_survey_paper}, we expanded the range further to study a variety of results) was created. Table~\ref{tab:sensor_preprocessing} shows the generated features per each sensor modality.

%Indirect Calorimeter (VO2) values equivalent to zero were updated to null as suggested in the WEEE \cite{weee} study.

% $$
% 1\;MET = \frac{3.5\;ml\:O_2}{kg * min} \eqno{(1)}
% $$

\begin{table}[t]
    \centering
    \caption{sensor pre-processing details}
    \label{tab:sensor_preprocessing}
    \begin{tabular}{p{0.8cm} l p{1.2cm}}
         \rowcolor[HTML]{EDEDED}
         \textbf{Sensors} &
         \textbf{Derived Features} & 
         \textbf{Libraries}
         \\

        \arrayrulecolor{Gray}

        ACC, GYRO &
        \makecell[l]{sum\_values, median, mean, length, \\standard\_deviation, variance, \\root\_mean\_square, maximum,\\ absolute\_maximum, minimum} &
        \makecell[l]{tsfresh \cite{tsfresh}}
        \\

        \arrayrulecolor{Gray}
        \hline
        
        PPG &
        \makecell[l]{bpm, ibi, sdnn, sdsd, rmssd, pnn20, \\pnn50, hr\_mad, sd1, sd2, s, sd1/sd2, \\breathingrate} &
        \makecell[l]{HeartPy \cite{heartpy}}
        \\

        \arrayrulecolor{Gray}
        \hline

        EDA &
        \makecell[l]{eda\_raw, eda\_clean, eda\_tonic, \\eda\_phasic, scr\_onsets, scr\_peaks, \\scr\_height, scr\_amplitude, \\scr\_rise\_time, scr\_recovery} & 
        \makecell[l]{NeuroKit \cite{neurokit}} 
        \\

        \arrayrulecolor{Gray}
        \hline
        
        EEG &
        \makecell[l]{skew, kurtosis, sample\_entropy, \\power\_spectral\_density based features, \\wavelet based features, \\hjorth\_mobility based features} &
        \makecell[l]{tsfresh \cite{tsfresh}, \\pywt \cite{pywt_Lee2019}, \\pyentrp}
        \\
        
        \arrayrulecolor{Gray}
        \hline
    \end{tabular}
\end{table}

%Furthermore, using tsfresh \cite{tsfresh}, statistical features such as median, mean, standard deviation, minimum, maximum values were calculated per each derived feature of PPG and EDA sensors.

%\subsection{Statistical Analysis}
%First, we separated the dataset into two classes to analyze each feature's significance in estimating MET values. \lakmal{Class 1 (Low-MET) contained data where the MET values were less than the median, MET value across the dataset, and Class 2 (High-MET) contained data where the MET values were greater than or equal to the median of MET values. Then we calculated the Cohen's-d (with 95\% confidence interval) \cite{rice2005comparing} \footnote{Rule of thumb to interpret Cohen's-d: 0.2 - Low Effect Size, 0.5 - Medium Effect Size, 0.8 - High Effect Size.} to determine the statistical significance of each feature in separating the MET values into either High-MET or Low-MET.}

%, catboost:unbiased_boosting

\paragraph{Energy Expenditure Estimation with Regression}
After pre-processing, we used principle component analysis with variance preservation of 99\% for dimensionality reduction. We tried out several models such as XGBoost \cite{xgboost:2016:XST:2939672.2939785}, CatBoost \cite{catboost:gradient_boosting}, LightGBM \cite{lightgbm:ke2017}, and some other models used in EEE literature \cite{ee_survey_paper} such as Linear Regression (LR) and Support Vector Machine (SVM). For SVM and boosting algorithms we performed hyperparameter tuning using Optuna \cite{optuna_2019}, a hyperparameter optimization framework. To avoid bias in the results, we performed leave-one-out cross-validation. For performance evaluation, we used Mean Absolute Error (MAE) and Root Mean Square Error (RMSE).  

% we performed nested cross-validation with group k-fold with k = 15 (grouping by removing a user) for the outer loop and a group k-fold with k = 14 (again grouping by removing a user) for the inner-loop. For performance evaluation, we used Mean Absolute Error (MAE), Root Mean Square Error (RMSE).
\section{RESULTS}

\begin{figure*}
    \centering
    \captionsetup{justification=centering, width=0.85\linewidth}
    \includegraphics[scale=0.165]{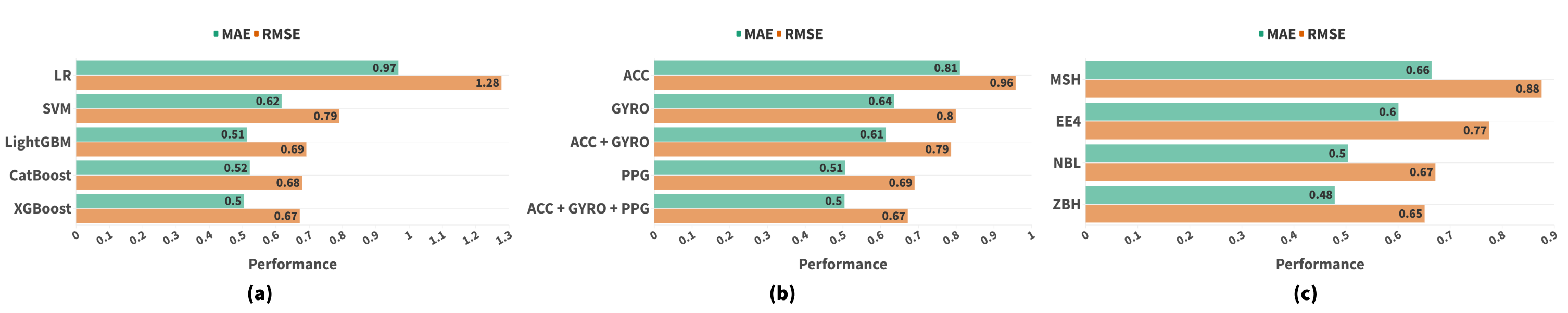}
    \caption{Performance Evaluation for (a) Earables, (b) Sensor Combinations, and (c) Devices.}
    \label{fig:combined_performance}

\end{figure*}

\paragraph{RQ1: EEE with Multimodal Earable Sensors.} We studied the earable's capability of measuring EE in many aspects. First, we evaluated the earable's EEE performance with five models trained with data from all three sensors for all six-time windows. We identified that the accuracy of the 6-second time window is relatively better for all models. As shown in Fig. \ref{fig:combined_performance}a, the Linear Regression model has the lowest results {(MAE of 0.97)} when compared to the other models. XGBoost, LightGBM, and CatBoost captured the non-linearity nature of the problem comparably better {(with MAEs of 0.5, 0.51 and 0.52 respectively)}, but the XGBoost has slightly outperformed LightGBM and CatBoost. Moreover, the XGBoost model provided a R\textsuperscript{2} score of 0.53, which is better compared to past EEE studies (LeBoeuf et al. \cite{leboeuf} were only able to obtained a R\textsuperscript{2} score of 0.36). Fig. \ref{fig:combined_performance}b shows performances of the sensor combinations that have been used for EEE using the XGBoost model. We got lower results {(MAE of 0.81)} for the model only with the accelerometer sensor and relatively higher results {(MAE of 0.51)} for the model only with the PPG sensor. However, the model containing all three sensors slightly performed better with an MAE of 0.5.

\paragraph{RQ2: Comparison to Other Wearables} First, we selected four devices (other than the Indirect Calorimeter) that were used in EEE. Fig. \ref{fig:combined_performance}c shows a comparison of performance between those devices for EEE. Prior work has shown that the chest is the most suitable body location for EEE, while the wrist is the worst \cite{ee_survey_paper}. Furthermore, apart from the dataset used in this study, to the best of our knowledge, no other studies used Muse S Headband for EEE. Further, according to Álvarez-García et al. \cite{ee_survey_paper}, there are no impactful studies conducted using the head as a body location for EEE. With the Muse S Headband performing the worst {(MAE of 0.66)} according to our study, this could indicate that the head might be worse than the wrist as a body location for EEE. Furthermore, Nokia Bell Labs Earbuds obtained a decent performance {(MAE of 0.5)} for EEE, and this could be due to its stability on the human body \cite{roddiger} while performing activities and heart rate features (i.e., bpm, ibi, sdnn, etc.) being informative for the inference.
\section{DISCUSSION}

\paragraph{Implications}
Our work has practical and theoretical implications. From a theoretical standpoint, our results add empirical evidence to the body of research on EEE using wearable devices. We showed that similar to other commercial-grade wearable devices \cite{o2020well}, earables can also be used for EEE with comparable performance.  The sensor combination results obtained by answering \textbf{RQ1} show that PPG sensor has performed better at measuring EE compared to Accelerometer sensor. Although in the past literature Accelerometer sensors have performed well in EEE, as to our knowledge there is only one study \cite{leboeuf} conducted with both Accelerometer and PPG sensors. However, there are no studies done on separate sensor modalities. Therefore, our research shows that in-ear PPG measurements could provide a more accurate EEE compared to in-ear Accelerometer measurements. With the Muse S Headband performing the worst according to our study, this could indicate that head might be worse than wrist as a body location for EEE. Furthermore, according to the results of \textbf{RQ2}, head could be considered the worst body location for EEE, especially when more informative heart rate-related signals could not be captured from there. From a practical standpoint, our findings speak to both earbud consumers and manufacturers. Earable devices are uniquely positioned to offer "continuous" monitoring, and they can also verify that the device is still being worn by the user \cite{roddiger}. With the added capability of EEE, consumers of earable devices will benefit from continuous EEE compared to other wearable devices. Also, using the results obtained by answering \textbf{RQ2}, it shows that Nokia Bell Labs Earbuds have performed better at EEE compared to other devices (except Zephyr BioHarness chest-band) worn in different body positions. A significant increase in the revenue for Apple's AirPods (nearly 600\%) \cite{noauthor_40_2023}, highlights the growing popularity of earable devices. This trend suggests that the capability for EEE will be a valuable addition for manufacturers to incorporate into their product offerings. 

\paragraph{Limitations and Future Work}
Our work has several limitations that call for future research efforts. The dataset we used contains three main activity types, each of which was conducted under two different intensity levels. However, more activity types would be required to generalize the model performance for real-world activities performed usually by humans. According to the Compendium of Physical Activities \cite{compendium}, there are 821 activities that can be categorized under 21 activity types. Therefore, future studies could be conducted with a greater number of activities in order to replicate real-life scenarios as much as possible. Further, the dataset we used was collected in a laboratory environment. Although it is more accurate and easy to conduct the experiment in such a setup, whether the data can be generalized to real-life situations remains unknown. In addition, in the study, activities were conducted in pre-defined time windows, and intensity levels were increased gradually. However, in a real-life setup, activities could change rapidly and conducted in different intensity orders. Therefore, future studies could consider creating more experiments in real-world setups. Finally, our study also provides insight into further exploring the head as a body location for EEE.
\section{CONCLUSION}

Although there are gold standards for measuring EE, they contain certain limitations, and wearable devices have already shown EEE capability with reasonable performance. The capability of accurately measuring EE using an earable device was presented in this study. We found that earables perform reasonably when data were captured in different activity levels and types. We also found that using PPG features provided better performance compared to accelerometers or gyroscopes. However, the feature combination worked the best with an MAE of 0.5 MET. We also found that earable performed decently compared to the chest-belt and outperformed the wristband and the headband. 

\bibliographystyle{IEEEtran}
\bibliography{src/citations}{}

\end{document}